\documentclass[amssymb,aps,showpacs,amsmath,pra]{revtex4}

\usepackage{graphicx}
                   
\usepackage{latexsym}

\newcommand{\be}{\begin{equation}}
\newcommand{\ee}{\end{equation}}

\newcommand{\bex}{\begin{eqnarray}}
\newcommand{\eex}{\end{eqnarray}}
\newcommand{\bmin}{\begin{center}\begin{minipage}{460pt}}
\newcommand{\emin}{\end{minipage}\end{center}}
\begin{document}

\title{Quantum Bit Commitment with a Composite Evidence}
\author{R. Srikanth}
\email{srik@rri.res.in}
\affiliation{Optics Group,
Raman Research Institute Bangalore- 560~080, Karnataka, India.}


\begin{abstract}
Entanglement-based attacks, which are subtle and powerful, are usually
believed to render quantum bit  commitment insecure. We point out that
the  no-go  argument  leading  to  this view  implicitly  assumes  the
evidence-of-commitment  to be  a monolithic  quantum system.  We argue
that  more  general evidence  structures,  allowing  for a  composite,
hybrid (classical-quantum) evidence,  conduce to improved security. In
particular,  we  present  and  prove  the security  of  the  following
protocol:  Bob  sends Alice  an  anonymous  state.  She inscribes  her
commitment $b$  by measuring  part of  it in the  + (for  $b =  0$) or
$\times$  (for  $b=1$)  basis.   She  then  communicates  to  him  the
(classical) measurement outcome  $R_x$ and the part-measured anonymous
state  interpolated  into  other,  randomly  prepared  qubits  as  her
evidence-of-commitment.
\end{abstract}
\maketitle

\section{Introduction}
Quantum cryptography draws its power from the very
principles of quantum mechanics, rather than, as with classical cryptography,
unproven assumptions about the hardness of certain computations
\cite{wie83}.  It has found its dominant application in   
quantum key distribution \cite{bb84}, which is provably secure 
(cf. \cite{got00} and references therein) and also implementable with current
technology \cite{gis01}. 
The ``post-cold war" applications of cryptography concern such tasks
as quantum coin tossing \cite{brcr91,bcjl,ard95,aha00,mso9,amb01},
quantum gambling \cite{gold},
quantum oblivious mutual identification \cite{crsl95}, quantum 
oblivious transfer \cite{cre94,ard95a} and two-party secure computations 
\cite{hoi97}, essentially concerned with secure processing
of the private information of mistrustful parties to reach a public decision. These are
closely related to quantum bit commitment (QBC) \cite{bb84,bcjl}, a 
quantum cryptographic
primitive for secure information processing. In a concrete if naive realization
of bit commitment, the committer (called Alice) writes 0 or 1 on a note, puts 
it into a safe, which she hands over to the acceptor (called Bob) as her evidence
of commitment. Upon Bob choosing to enter the
transaction, she gives him the key to the safe. The main point is that
Alice should not be able to cheat by changing her mind after handing Bob the safe, nor
should Bob be able to cheat by finding
out about Alice's decision until after she gives him the key. A secure bit commitment
is one which is (at least, exponentially in some security parameter) 
binding on Alice {\em and} 
unconditionally concealing (of her commitment) from Bob and thus prevents
either party from cheating. 

However, it is generally agreed that secure QBC is impossible \cite{may,loc} because
of the possibility of an entanglement-based \cite{epr} attack by Alice.
Here a dishonest Alice sends as evidence of her commitment $b \in \{0,1\}$ towards Bob
photons in entangled states instead of ones in a definite polarization
state. The ensemble $\chi_b$ and the corresponding density matrix $\rho_b$ of possible
states representing commit bit $b$
should satisfy $\rho_0 = \rho_1$ in order to be indistinguishable to Bob. Then,
according to the Gisin-Hughston-Jozsa-Wootters (GHJW) theorem \cite{hug93}, a
purification of $\chi_0$ can be rotated remotely to that of $\chi_1$.
Therefore, by delaying her measurement until after unveiling, Alice can cheat by
unveiling a state in $\chi_0$ or $\chi_1$. This powerful
argument forms the basis of the no-go argument for QBC.
Related quantum two-party secure computation protocols \cite{hoi97},
simultaneously secure against both Alice and Bob are also believed to be
impossible, except in a relativistic scenario \cite{ken00bc},
though a trade-off is permitted \cite{spe01}.
Yet, arguments have been put forward pointing out that 
the no-go result for QBC does not cover all possible QBC scenarios
\cite{yuenqbc,yuenqkd,yuennonogo,he,yuenny,arian}.
This discrepancy is partly explained by the difficulty of
characterizing all possible scenarios.
An important step towards remedying this situation is 
a recent classification of protocols \cite{arian} that
includes anonymous-state based protocols, introduced by Yuen
\cite{yuenqbc}, and thus allows for more general schemes than the Yao model
\cite{yao95} assumed in the no-go argument.

In a typical QBC scheme with BB84 encoding \cite{bb84}, 
qubits (photons) come in four possible preparations:
in the rectilinear basis (denoted +),
with horizontal (signifying 0) or vertical (signifying 1) 
polarization; else, in the diagonal basis \mbox{(denoted
$\times$)} with polarization oriented at $45^{\circ}$ (signifying 0) or 
$135^{\circ}$ (signifying 1). During the commitment phase,
Bob receives from Alice an evidence-of-commitment, whose state she unveils along
with $b$ in the unveiling phase. The new protocol we present differs from this
pattern in three significant ways: inclusion of Bob's anonymous state, the classical
component of the evidence, and Alice's use of decoy qubits. The first two features
are indispensable.

\section{The new protocol\label{tnp}}

We present herebelow a new protocol, denoted $\mathbb{P}$,
the proof of whose security against entanglement-based and other attacks
is given thereafter. For simplicity,
it assumes a noiseless channel but can easily be extended to the noisy case. 
Let $m, n, p, q$ be four pre-agreed security parameters such that
$1 \ll m \ll n \ll p, q $. The complete honest protocol consists of 
three phases: (1) pre-commitment
phase, (2) commitment phase, (3) unveiling phase. The intervening period between
the commitment- and unveiling-phase, sometimes called the holding phase,
can be arbitrarily long. A note on notation: $[y,z]_b$ represents a function that
takes value $y$ ($z$), when $b=0$ ($b=1$). 
\begin{enumerate} 
\item {\bf Pre-commitment phase}:
\begin{enumerate}
\item Bob chooses two random, unknown-to-Alice, $p$-bit strings 
$R_B \in \{0,1\}^{p}$ and $\eta \in \{+,\times\}^{p}$.
He prepares the pure, separable $p$-qubit state
\mbox{$|R_B\rangle_{\eta} = 
|R_B(1)\rangle_{\eta(1)}\otimes\cdots\otimes|R_B(p)\rangle_{\eta(p)}$}.
\item He sends the anonymous state $|R_B\rangle_{\eta}$ to Alice over a quantum channel. 
\end{enumerate}
\item {\bf Commitment phase}:
\begin{enumerate}
\item Test for random mixing: Alice randomly chooses $(p-n)/2$ qubits from 
$|R_B\rangle_{\eta}$ and measures them in
$+$ basis and checks that she obtains almost equal 0 and 1 outcomes. She repeats
the same for $\times$ basis. (If even one of the checks fails, she aborts 
the protocol run,
convinced that Bob biased the system he sent her.) The measured qubits are
discarded. We denote by
$|\tilde{R}_B\rangle_{\tilde{\eta}}$ the state of the $n$ 
undiscarded qubits remaining with her, and
by $P$ the $p$-bit string specifying the positions of the undiscarded qubits.
The tilde (over $R$ and $\eta$)  denotes restriction to the undiscarded qubits.
During discarding, the ordering of 
the surviving qubits is preserved \cite{order}.
\item Measurement on $m$ undiscarded qubits: She generates a random
$n$-bit string $x \in \{0,1\}^n$ of Hamming weight $m$ (ie., number of 1's is $m$).
Of the surviving qubits, 
a qubit $i$ for which $x(i)=1$ ($x(i)=0$) is called `marked' (`unmarked').
To commit to $b=0$ ($b=1$), she measures the $m$ marked
qubits in the $+$ ($\times$) basis. The $m$-bit measurement outcome is denoted $R_x$.
\item Introduction of decoy qubits: Alice chooses a random $q$-bit string $Q$ 
of Hamming weight $n$ ($\ll q$). She creates a $q$-qubit state 
as follows: in slot $i$, if $Q(i) = 0$, she inserts a decoy qubit in an arbitrary
state, else she inserts an undiscarded qubit, in sequential order.
The resulting state is denoted $|R_A\rangle_{\theta}$, where $R_A \in \{0,1\}^q$
and $\theta \in \{+,\times\}^q$ \cite{simp}.
\item Evidence communication: She communicates to Bob the triple
($P$, $R_x, |R_A\rangle_{\theta}$) 
as her evidence of commitment: $P$ and $R_x$ over a classical 
channel, $|R_A\rangle_{\theta}$ over a quantum channel.
\end{enumerate} 
\item {\bf Unveiling phase}:
\begin{enumerate}
\item Alice announces $Q$, $b$ and $x$. 
\item Using $Q$, Bob locates and removes the decoy qubits. Using $x$,
he locates the $n-m$ unmarked (undiscarded) qubits, and verifies that 
he recovers $|R_B^{\prime}\rangle_{{\eta}^{\prime}}$, where the prime denotes restriction
to unmarked qubits. 
\item Measuring all marked qubits in 
the basis $[+,\times]_b$, he verifies that he obtains outcome $R_x$.
\item On the marked qubits, if $[+,\times]_b = \eta^{\prime\prime}(i)$, 
he checks that $|R_x(i)\rangle_{[+,\times]_b} = 
|R_B^{\prime\prime}(i)\rangle_{\eta^{\prime\prime}(i)}$, where the double prime
denotes restriction to marked qubits. 
\end{enumerate}  
\end{enumerate} 
The basic intuition behind the protocol can be summarized as follows: Alice encodes
$b$ by measuring part of the anonymous state sent by Bob in the + (for $b=0$) 
or $\times$ (for $b=1$) basis. She then communicates the
(classical) outcome $R_x$ and the part-measured anonymous state to him
as evidence. Announcing $R_x$
almost entirely deprives her of the freedom to depart from honest execution
because the measured qubits were in unknown-to-her states prepared by Bob. 
On the other hand, Bob is unable to exploit his prior knowledge of $|R_B\rangle_{\eta}$
and received knowledge of $R_x$ on account of her insertion of decoy qubits, which
serve as ``junk" information, preventing him identifying the ``signal" 
qubits. A more detailed proof is given herebelow.

\subsection{Security against Bob}
Before the unveiling, because of 
his prior knowledge of $|R_B\rangle_{\eta}$ and  
Alice's announcement of $R_x$ in step 2(d), 
Bob knows that an $m$-qubit subset of quantum evidence received from Alice
is in the state $|R_x\rangle_{\kappa}$, where $\kappa$ is all
+ or all $\times$ and another, \mbox{$(n-m)$-qubit} subset in the state originally
prepared by him. If Bob knew which $n$ qubits were non-decoy qubits, 
he could measure them all in the $\tilde{\eta}$ basis, and based on the departure
from $|\tilde{R}_B\rangle_{\tilde{\eta}}$, he could with high probability
($= 1-(1/2)^{m/2}$) determine $b$. And if he knew which $m$ qubits were marked, 
he could measure them all either in + or $\times$ basis,  and based on the departure
of outcomes from $R_x$, he could similarly with high probability
($= 1-(1/2)^{m}$) determine $b$. But in both these he is thwarted by the classical 
combinatorial uncertainty arising from
the exponentially large number of ways in which Alice's 
$n$ non-decoy qubits could be scattered among the $q$ return qubits:
for $n \ll q$, $Q$ could have been chosen in about $(q/n)2^n$ ways \cite{howmany}. 
So the probability that knowledge of $R_x$ and $|R_B\rangle_{\eta}$
will help him to correctly infer $b$  
is exponentially small in $n$. For the same reason, he is unable to employ  
an entanglement-based attack wherein he sends entangled qubits, and
by pairing up Alice's return qubits with their hidden entangled counterparts,
tries to identify $b$ by comparing correlated measurements in 
some fixed basis. 
As a result, from Bob's viewpoint the quantum system sent by Alice is
exponentially close to the maximally uncertain state $2^{-q/2}\hat{I}^{\otimes q}$.
Finally, we note that if he biases $|R_B\rangle_{\eta}$ 
by sending all qubits in an identical
state (or with the distribution of 0's and 1's being basis dependent), 
Alice will most certainly detect this in step 2(a). 
This completes the proof of security against Bob. 

\subsection{Security against Alice} 
Her evidence consists of an auxiliary, two-part classical 
component ($P, R_x$) and a quantum component
($|R_A\rangle_{\theta}$). We will find that the origin of the security against Alice is that
the classical component of the evidence restricts what she can achieve by nonlocally influencing 
the state of the quantum evidence, thereby constraining her to play honestly.
A classical record has only one possible ensemble realization: it cannot be 
set in a superposition state. In particular, it does not permit two distinct
equivalent ensembles that can be rotated into each other in the
sense of the GHJW theorem. Therefore, no
entanglement-based attack based on $(P, R_x)$ is possible.
Another way to see this is that if Alice could
alter $(P, R_x)$ via entanglement, this would have led to superluminal signaling
\cite{ghi88,sri01}. Hence any possible attack on her part should be based on
a fixed $(P, R_x)$ after step 2(d). Further,
she is restricted by quantum no-cloning \cite{woo82} and
Heisenberg uncertainty from knowing about $|R_B\rangle_{\eta}$ 
and must operate within the confines of this ignorance.

From Alice's viewpoint, $|R_A\rangle_{\theta}$ factors into three
distinct sectors:  the $n-m$ unmarked qubits 
in an unknown-to-her state ($\in {\cal H}_{\nu}$, the unmarked subspace), 
the $m$ marked qubits in a known-to-her state ($\in {\cal H}_{\cal C}$, the coding subspace) 
and the $q-n$ decoy qubits in a known-to-her state ($\in {\cal H}_{\cal D}$, the decoy subspace). 
The state of the quantum evidence is given by
$|R_A\rangle_{\theta} = |\phi\rangle \otimes
|R_A^{\prime}\rangle_{\theta^{\prime}}
\otimes |R_x\rangle_{[+,\times]_b} \in {\cal H}_E \equiv {\cal H}_{\cal D}\otimes
{\cal H}_{\nu}\otimes{\cal H}_{\cal C}$. Now, she must leave the system ${\cal H}_{\nu}$ untouched
in order to pass step 3(b), where Bob checks that the unmarked qubits remain
unmeasured. Therefore, no attack may involve this sector. Further note that the
announcement of $P$ means that discarded qubits can never be unveiled as marked or
unmarked qubits. Therefore, Alice's any possible attacks should be
confined to the subspace ${\cal H}_{\cal D} \otimes {\cal H}_{\cal C}$.
To prove security, we need to show that no attacks exist that can pass Bob's
both checks in steps 3(c) and 3(d).
We first consider attacks {\em not} based on entanglement.

\paragraph{Attacks not based on entanglement:}
Suppose she executes step 2(b) honestly with $b=0$ and announces $R_x$ in step 2(d).
Her chances of dishonestly unveiling $b=1$ by unveiling $|R_x\rangle_+$
as $|R_x\rangle_{\times}$, and hence passing Bob's check in 3(c),
is exponentially small because the overlap
$_{\times}\langle R_x|R_x\rangle_+ = (1/2)^{m/2}$. 
Nor can she cheat by interchanging marked and decoy qubits: for while she can 
obviously unveil marked qubits as decoy qubits, the converse is not true.
For example, suppose she prepares
$|R_x\rangle_+$ honestly, but dishonestly prepares $m$ of her decoy
qubits in the state $|R_x\rangle_{\times}$. To unveil them
as her marked qubits will allow her to pass test 3(c). However, unveiling
the dishonest $b=1$ will almost certainly (with probability 
$(1 - 2^{-m})$) lead to her
being caught in step 3(d). Here we note that if $P$
were not part of evidence submitted in step 2(d), then she could have cheated 
simply by introducing suitable discarded qubits found in the state $|R_x\rangle_{\times}$ 
among the decoy qubits.

\paragraph{Mayers-Lo-Chau attack:}
Let us consider the prospects of an entanglement-based attack.
The main point here is that Alice should have made her honest measurement
before evidence communication step 2(d). If, without measuring, she announces
an arbitrary $R_x$, entangling and deferring her measurement, her chance of escaping 
step 3(c) is $2^{-m}$ (the probability to obtain measurement projection
$|R_x\rangle_{\kappa}\langle R_x|_{\kappa}$ where $\kappa$ is all $+$ or all $\times$). 
Next, we note that for the reason mentioned in the preceding paragraph, decoy qubits
cannot be passed off as marked qubits, because of step 3(d).
Therefore, any attempt at an entanglement-based attack 
must be restricted to sector ${\cal H}_{\cal C}$. 
But here, we observe that $|R_x\rangle_+\langle R_x|_+ \equiv \rho_0^{\cal C}
\ne |R_x\rangle_{\times}\langle R_x|_{\times} \equiv \rho_1^{\cal C}$, 
where the superscript ${\cal C}$ refers to 
${\cal H}_{\cal C}$. Thus, these two ensembles, which code for her two possible
commitments, conditioned on Bob's knowledge of $R_x$, are {\em inequivalent},
with the consequence that a purification of $\rho_0^{\cal C}$ cannot be remotely
rotated into one of $\rho_1^{\cal C}$ in the sense of the GHJW theorem. 
It follows that an entanglement-based attack of the type envisaged by
the no-go argument is not possible. 
Another way to see this is that if it were possible, it would lead to superluminal signaling. 
At the same time, the fact that $\rho^{\cal C}_0 \ne \rho^{\cal C}_1$
does not lead to distinguishability with respect to Bob, because
the state $\rho_b$ of the evidence {\em as a whole} indeed satisfies $\rho_0 = \rho_1$. 

This elucidates how the composite, hybrid structure of the evidence in the protocol 
is essential to evade the no-go argument, which
implicitly assumes a monolithic (i.e., non-composite), purely quantum evidence, where 
${\cal H}_{\cal C} = {\cal H}_{E}$, i.e., the coding space is all of the quantum evidence.
Of course, in such case, the requirement $\rho_0=\rho_1$ on ${\cal H}_E$ would imply
$\rho_0^{\cal C}=\rho_1^{\cal C}$, from which the Mayers-Lo-Chau attack follows. 
The desired twin features of
security against Bob and that against Alice refer to {\em different} state spaces
(${\cal H}_{E}$ and ${\cal H}_{\cal C}$, respectively), that is,
state indistinguishability to ${\cal H}_E$ (for security against Bob) and state
inequivalence to ${\cal H}_{\cal C}$ (for security against Alice).
As a result, guaranteeing the former does not imply violation of the latter.

\paragraph{Weaker entanglement-based attacks:}
Now that we have demonstrated security against the standard entanglement-based attack,
we turn our attention to weaker attacks.
The Mayers-Lo-Chau attack, where applicable, is characterized by:
\begin{equation}
\label{Eq:mlc}
p^B_{\rm cheat} \rightarrow 1/2 ~\Longrightarrow~ p^A_{\rm unveil}(b)
\rightarrow 1,
\end{equation}
where $p^A_{\rm unveil}(b)$ is the probability that Alice
can successfully unveil $b$, and $p^B_{\rm cheat}$ is
Bob's cheating probability. The minimum value (1/2) of $p^B_{\rm cheat}$ 
corresponds to a plain guessing chance for Bob. Eq. (\ref{Eq:mlc}) says that if the
probability for Bob to cheat approaches the guessing value, then that for Alice
to unveil any value of $b$  is 1. 
A more general attack we can envisage is one characterized by:
\begin{equation}
\label{Eq:weak}
p^B_{\rm cheat} \rightarrow 1/2 ~\Longrightarrow~ p^A_{\rm unveil}(b)
\rightarrow \beta(b),
\end{equation}
where $\beta(b)$ does not vanish asymptotically as a function of any security parameter.
It would seem reasonable to require that at least one of the two $\beta(b)$'s
should be 1, meaning she can cheat with complete certainty for 
at least one value of $b$.  
Yet, in this general criterion, we won't even demand that. For example,
it could be that $\beta(0) = \beta(1) \approx 0.5$. 
Now, why would Alice want to launch such a weak entanglement-based attack, 
where she is not quite sure what $b$ she will have to unveil? 
One possible scenario is that she is an unknown player, who loses nothing 
even if her dishonesty is detected. 
The main point is that where a weak attack exists,
$\lambda \equiv \min\{\beta(b)\} > 0$ 
(counting the Mayers-Lo-Chau attack as a special
case where $\lambda = 1$). We seek security even against this more general condition, such that
$\lambda = 0$ (asymptotically).
A protocol for which $0 < \lambda < 1$, which is secure against a standard entanglement-based
attack, but not against a weak attack, is called {\em weakly secure}. 
A protocol for which $\lambda = 0$ (asymptotically) is said to be strongly secure. 

An example for a weakly secure protocol, $\mathbb{P}^{\prime}$, is as follows: 
modify step 2(b) in the above protocol to one wherein Alice, instead of measuring
$m$ anonymous qubits,
prepares and interpolates an $m$-qubit state $|R_x\rangle_{[+,\times]_b}$ in the
marked positions among $n-m$ unmarked anonymous qubits. In step 2(c), she does
not introduce decoy qubits, but scrambles the $n$-qubit state in ${\cal H}_{\nu}
\otimes {\cal H}_{\cal C}$ according to permutation $\Pi$ 
in such a way that the relative ordering of the marked qubits among
themselves remains fixed. In 2(d), she sends $R_x$ and the system ${\cal H}_{\nu}
\otimes {\cal H}_{\cal C}$ as composite evidence. In step 3(a), she announces
$b, x, \Pi$. Obviously, step 3(d) is dropped. It may be verified that this protocol
is secure against Bob and against a standard entanglement-based attack by Alice
for the same reasons as $\mathbb{P}$.
However, it is only weakly secure because the following weak entanglement-based
attack exists: in step 2(c) Alice interpolates the second register in the state
\mbox{$|\zeta\rangle \equiv (1/\sqrt{2})(|0\rangle|R_x\rangle_+ + 
|1\rangle|R_x\rangle_{\times})$}. After the commit phase, she measures the first
register and unveils $b=0$ ($b=1$) if she finds 0 (1). She has 50\% chance of
cheating successfully. The protocol is characterized by $\lambda = 0.5 = \beta(0) =
\beta(1)$.

We now demonstrate that the protocol $\mathbb{P}$ is {\em strongly secure}.
To this end, one needs to consider if there is an 
optimal measurement \cite{eke94}
she can perform in 2(b) to obtain a value of $R_x$ that, with a
finite probability independent of $m$, is an outcome she could obtain in {\em both}
$+$ and $\times$ basis. If yes, she can create the state $|\zeta\rangle$ and
proceed as the attack on $\mathbb{P}^{\prime}$.
To see that this is exponentially impossible, we note
that there is {\em exactly one} bit-string $R_x$ consistent with being unveiled
in both $+$ and $\times$ basis \cite{whynot}. Further, before measurement in step
2(b), the $m$ marked qubits are, from Alice's viewpoint, maximally
uncertain, i.e., the density operator is given by $2^{-m}\hat{I}^{\otimes m}$. 
It follows that there is no general positive operator-valued measure \cite{nc,pre}
that can yield this special outcome with a probability greater than $2^{-m}$.
Hence the chance that she can obtain an $R_x$ that is a possible
valid outcome in both $+$ and $\times$ basis, and thus launch a weak attack using
a $|\zeta\rangle$ based on such $R_x$, is exponentially small in $m$.
This completes the proof of security against Alice's weak entanglement-based attack. 

\section{Conclusion}
In contrast to a no-go result like no-cloning \cite{woo82}, that for
QBC is more complicated to characterize. The reason is that whilst
cloning is a single well-defined physical process, QBC is a cryptographic task,
whose decomposition into individual processes can be model-dependent, with no obvious
indication of the `most general' model. For the
no-go argument to be truely universal, it must demonstrate that the model 
(denoted, say, $\mathbb{M}$) which it proves insecure is the most general allowed. 
In retrospect, some scope-restricting features of $\mathbb{M}$ are evident.
It appears that these features may have been assumed because 
$\mathbb{M}$ builds QBC implicitly on the basis of classical bit commitment adapted 
to include the Yao model for two-party protocols \cite{yao95}.

For one, modelling the evidence as a monolithic, (i.e., non-composite), 
purely quantum system that encodes $b$, it fails to explore 
more general evidence structures and their security implications. 
In our protocol, the evidence is composite and hybrid
(classical-quantum), with two classical ($R_x, P$) and three quantum 
(${\cal H}_E \equiv {\cal H}_{\cal D}\otimes {\cal H}_{\nu}\otimes{\cal H}_{\cal C}$)
parts. This composite and hybrid structure and the inter-relation between the 
constituent parts is critical to the protocol's security.
Second, the no-go argument assumes that any QBC scheme can be reduced before unveiling
to an equivalent scheme in which Alice and Bob share a mutually known entangled state. 
But this is incompatible with the use of the classical component $(P, R_x)$
of the evidence since a classical system cannot be put into a superposition, 
let alone be entangled.

Finally, in the light of our result, some issues concerning
no-go arguments in quantum ``mistrustful" cryptography
merit careful consideration. 
We can expect to secure coin tossing by building it on top of the proposed QBC 
\cite{qct}. Hence, it is important to clarify how our protocol
relates to no-go arguments for coin-tossing, 
e.g., the result for the lower bound on the number of sequential rounds for
a given bias in the quantum coin tossing, proposed
in Ref. \cite{amb01}, or the impossibility of ideal coin tossing, advanced in 
Ref. \cite{loc}.  It turns out 
our scheme does not directly affect them in that, being proposed after
the discovery of the Mayers-Lo-Chau attack, they were 
designed to be independent of the security of QBC.
But now these no-go results must be qualified as 
pertaining to coin tossing protocols {\em not} built on QBC \cite{comment}.
Similarly, we can also expect to secure quantum oblivious transfer 
\cite{yao95} by basing it on 
of the proposed QBC scheme. Kilian \cite{kil91} has shown that, in classical 
cryptography, oblivious transfer can be used to implement protocols such as
oblivious circuit transfer, which is related to secure two-party
computation. This chain of arguments re-injects hope into the realizability
of quantum versions of ``post-cold war" cryptographic tasks.

{\em Acknowledgements:} I am grateful to Prof. Guang-Ping He for valuable comments.
I am thankful to Prof. C. H. Bennett, Prof. G. Brassard, 
Prof. H.-K. Lo, Prof. J. Pasupathy and Prof. S. Raghavan for discussions. 
This work was partially supported by the DRDO project 510 01PS-00356.


\begin{thebibliography}{100}
\bibitem{order} To be precise: if $j_f, k_f$ are the indexes of two surviving qubits in
$|\tilde{R}_B\rangle_{\tilde{\eta}}$ and $j_i, k_i$ their respective initial
indexes in $|R_B\rangle_{\eta}$, then $j_i < k_i \Longleftrightarrow j_f < k_f$.
\bibitem{simp} A practical simplification here is that the decoy qubits may
prepared arbitrarily, and not necessarily in BB84 states.
\bibitem{howmany} 
The problem is equivalent to the number of combinations for placing $n$ items among
$q$ slots, which is $\left(\mathop{~}^{q}_{n}\right) \sim (q/n)2^n$,
where we have used Stirling's formula, and the fact that $n\ll q$.
\bibitem{whynot} Let us represent the state of the marked qubits before Alice's 
measurement by $|R_B^{\prime\prime}\rangle_{\eta^{\prime\prime}} \equiv 
|r\rangle_+|s\rangle_{\times}$, where $0 \le r, s \le 2^{m}-1$
and we have collected the qubits prepared in a fixed basis separately for 
convenience. The number of qubits in $|r\rangle_+$ ($|s\rangle_{\times}$) is denoted
$n(r)$ ($n(s)$).  We denote the first $n(r)$ (the remaining $n(s)$) bits
of $R_x$ by $R_x^{\uparrow}$ ($R_x^{\downarrow}$). To be consistent with a $+$ 
($\times$) basis measurement, $R_x^{\uparrow}$ ($R_x^{\downarrow}$) should
have the value $r$ ($s$). Therefore there is exactly one string, namely 
$R_x = r\oplus s$, that is consistent with being unveiled in $+$ or $\times$ basis.
\bibitem{qct} According to this well known result: Alice commits bit $a$
towards Bob, who announces bit $b$. Alice unveils $a$. Both consider that the
outcome of the coin toss is the random bit $a \oplus b$.
\bibitem{comment} Because coin tossing is
weaker than bit commitment \cite{ken00ct}, in principle one can try to build a
secure coin tossing (1) on top of bit commitment, or (2) independent of it.
Chronologically, Refs. \cite{loc,amb01} appeared after the discovery of
the Mayers-Lo-Chau attack on bit commitment. So they presume the impossibility
of coin tossing of type (1), and in fact study only aspect (2). Thus, strictly
speaking, no-go arguments for coin tossing derived therein refer only to type
(2) coin tossing. For example, the coin tossing template used in Ref. 
\cite{amb01}
for deriving the lower bound result reported therein employs a joint system
shared by Alice and Bob, subjected to a number of rounds of quantum
communication and unitary operations, as well as conditions like Lemma 11
and Lemma 12 of Ref. \cite{amb01}. 
None of these is a necessary ingredient in (a coin tossing
based on) QBC, where the basic protocol aims
not to share a random bit, but to securely and bindingly commit a bit.
\end{thebibliography}
\end{document}